# THE JEFFERSON LAB 1 KW IR FEL

D. Douglas, for the IR Demonstration FEL Project Team,
Thomas Jefferson National Accelerator Facility, Newport News, VA 23606, USA

*Abstract*

The Jefferson Lab (JLab) IR Demo Free Electron Laser (FEL) has completed commissioning and is initiating user service. The FEL — a high repetition rate, low extraction efficiency wiggler-driven optical cavity resonator — produces over 1 kW of tuneable light on intervals in a 3–6 μm wavelength range. It is driven by a 35–48 MeV, 5 mA superconducting RF (SRF) based energy-recovering continuous wave (CW) electron linac.

The driver accelerator meets requirements imposed by low energy, high current, and a demand for stringent beam control at the wiggler and during energy recovery. These constraints are driven by the need for six-dimensional phase space management, the existence of deleterious collective phenomena (space charge, wakefields, beam break-up, and coherent synchrotron radiation), and interactions between the FEL and the accelerator RF. We will detail the system design, relate commissioning highlights, and discuss present performance.

## 1 SYSTEM REQUIREMENTS

The FEL is a wiggler-driven laser with an 8 m long optical cavity resonator [1]. It uses moderate gain and output coupling, low extraction efficiency and micropulse energy, and high repetition rate to avoid high single bunch charge while producing high average power. This paradigm leads to the use of SRF technology, allowing CW operation, and motivates use of energy recovery to alleviate RF system demands.

The system architecture thus imposes two requirements on the driver accelerator:
- delivery to the wiggler of an electron beam with properties suitable for the FEL interaction, and
- recovery of the drive beam energy after the FEL.

The first requirement reflects the needs of the FEL system itself. Optimized beam parameters are given in Table 1. We note the nominal FEL extraction efficiency produced with these parameters is >½%. The micropulse energy is modest; high output power is achieved through the use of very high repetition rate (20[th] subharmonic of the RF fundamental) and CW operation.

The energy recovery requirement reduces RF system demands (both installed klystron power and RF window tolerances), cost, and radiation effects by decelerating the beam after the FEL so as to drive the RF cavities. As the full energy spread after the wiggler exceeds 5%, this creates a need for a large acceptance transport system.

Table 1: Optimized system parameters

| | |
|---|---|
| Beam energy at wiggler | ~40 MeV |
| Beam current | 5 mA |
| Single bunch charge | 60 pC |
| Bunch repetition rate | 74.85 MHz |
| Normalized emittance | 13 mm-mrad |
| RMS bunch length at wiggler | ~½ psec |
| Peak current | 60 A |
| FEL extraction efficiency | >½% |
| δp/p   rms, before wiggler | ¼ % |
|           full, after wiggler | 5% |
| CW FEL power | >1 kW |

## 2 SYSTEM CONFIGURATION

### 2.1 Overview

The above system requirements couple to many phenomena and constraints. Phase space requirements at the FEL demand transverse matching and longitudinal phase space management during acceleration and transport to the wiggler. Similarly, the machine must provide adequate transverse beam size control while managing the large longitudinal phase space. Such transport and conditioning of the beam must be performed in the presence of a number of potential collective effects driven by the high current and low energy. To avoid space-charge-driven beam quality degradation, moderately high injection energy is needed [2]. Beam break-up (BBU) and other impedance-driven instabilities must be avoided [3]. Coherent synchrotron radiation (CSR) must be managed to preserve beam emittance [4]. RF stability must be assured, particularly in transient regimes such as FEL turn-on and initiation of energy recovery [5].

Figure 1 illustrates the system concept, which has successfully addressed these issues. The schematic shows the 10 MeV injector, a single eight-cavity Jefferson Lab cryomodule accelerating to ~40 MeV, an FEL insertion, and energy recovery transport from wiggler through module to a beam dump. All acceleration is performed using standard CEBAF 1.497 GHz five-cell cavities. A summary of the function and performance of each section will now be provided.

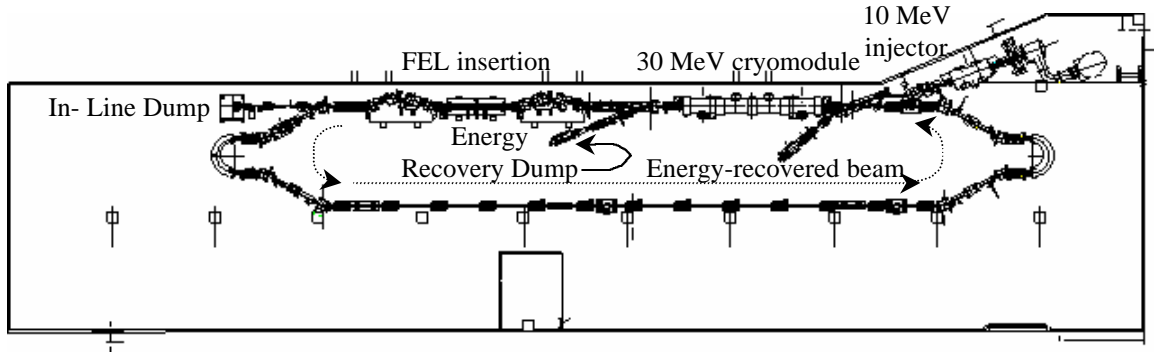

Figure 1: Jefferson Lab 1 kW IR FEL. The machine is shown in the facility vault.

## 2.2 Injector

The electron source is a DC photocathode gun nominally producing 60 pC bunches at 320 keV with repetition rates of up to 75 MHz [6]. Immediately following the gun a room-temperature buncher compresses the initial electron pulse, which is then captured by a two-cavity CEBAF cryounit and accelerated to 10 MeV. A four-quad telescope matches beam envelopes to the linac acceptance across a three-bend "W" achromat. RF component phases are adjusted to produce, in concert with the injection line momentum compaction, a long (~3 psec rms) low relative momentum spread (~0.1% rms) bunch at the entrance of the linac.

Injected beam quality depends on gun operating voltage and charge per bunch; typical normalized emittances for 320 kV operation are of order 5–10 mm-mrad [7].

## 2.3 Linac

The linac accelerates the injected beam from 10 MeV to 35–48 MeV using a single high gradient eight-cavity Jefferson Lab cryomodule. By accelerating $8°$ off crest, a phase/energy correlation is imposed on the longitudinal phase space; this is used downstream for bunch compression. The RF cavities also provide transverse focussing, assisting in beam envelope management.

Immediately after the cryomodule, a small dipole is used to separate the accelerated and energy recovered beams. The low energy beam is directed to a dump; the effect of this bend on the full energy beam is corrected by a subsequent pair of small bends.

## 2.4 FEL Insertion

The FEL is located immediately beyond the linac. As this is prior to recirculation bending, it avoids potential CSR degradation of beam quality and allowed a low power "straight ahead" operational (non-energy recovering) mode before the recirculator was fully installed; this remains a useful diagnostic configuration. A quadrupole telescope (two triplets) matches beam envelopes from module to wiggler. An achromatic four-dipole chicane between the triplets separates optical cavity and electron beam components while compressing the bunch length. The chicane geometry is constrained by the tolerable momentum compaction. Larger chicanes provide more space but lead to higher compactions with more time of flight jitter; to maintain FEL pulse/drive beam synchronism, the chicane $M_{56}$ is restricted to –0.3 m.

The match from module to wiggler, by virtue of RF focussing, depends on linac energy gain. It is therefore adjusted operationally to compensate for gross (several MeV) energy changes. After the wiggler, the electron beam (full momentum spread > 5%) is matched to the recirculation transport using a second quad telescope. This avoids beam envelope mismatch, large spot sizes, aggravated optical aberrations, error sensitivities, and potential beam loss. As in the linac to wiggler transport, a dipole chicane embedded in the telescope moves the electron beam off the optical cavity axis; this chicane also lengthens the electron bunch, reducing peak currents and alleviating potential wakefield and CSR effects. Simulations and experience with the machine indicate that space charge effects are not significant above ~25 MeV [8]; analysis of system performance and operational tuning is therefore possible using single-particle transport models.

## 2.5 Recirculator/Energy Recovery Transport

Following the FEL insertion, the electron beam (full momentum spread > 5%) is transported through a recirculation arc to the linac for energy recovery. This recirculator provides both transverse beam confinement and longitudinal phase space conditioning. Bending is provided by achromatic and nominally isochronous end loops based on an MIT-Bates design [9]. Dipole parameters (bend and edge angles) and drift lengths are set to provide $M_{56}=0$ from wiggler to reinjection point, and, across each end loop, betatron stable motion in the horizontal plane (with a tune of 5/4) and imaging transport vertically (-I transfer matrix). The end loops are joined by six $90°$ FODO cells; with the end loop phase advances and reflective symmetry across the backleg, this suppresses aberrations over the full arc.

Beam path length through the recirculator is adjusted using steering dipoles adjacent to the large 180° dipoles and is used to set the phase of the energy-recovered beam with respect to the module RF fields. Each end loop has four trim quads and four sextupoles for dispersion and compaction control. A single family each of quadrupoles and sextupoles (adjacent to the 180° bends) is used to modify the linear and quadratic momentum compactions from wiggler to reinjection, so as to compensate the slope and curvature of the RF waveform during energy recovery. This allows simultaneous recovery of RF power from the electron beam and compression of the beam energy spread at the dump.

The second end-loop delivers the longitudinally conditioned beam to the linac axis, where it is betatron matched to the cyromodule acceptance using a four-quad telescope, and merged with the injected beam using a small achromatic three-dipole chicane.

## 2.6 Longitudinal Matching Scenario

Key to the operation of this device is the use of bunch length compression (to create high peak current for FEL gain) and energy recovery (to provide RF power required for acceleration of high average currents) [10]. Figure 2 presents a schematic of the longitudinal matching scenario employed in the system. The individual phase-energy plots indicate the orientation of the longitudinal phase space at key locations around the machine.

The injector provides a long, upright, small momentum spread bunch (3 psec rms × 30 keV rms), which is accelerated off-crest in the linac. This imposes a phase-energy correlation, generating ~¼% momentum spread — about 100 keV at 40 MeV — over one rms bunch length. The momentum compaction of the chicane upstream of the wiggler rotates this slewed phase space upright, generating a short bunch (0.4 psec rms) at the wiggler.

The FEL interaction does not affect bunch length, but does generate a large full momentum spread. This is evident in Figure 3, which shows the beam at a dispersed point ($\eta$=0.4 m) in the chicane immediately downstream of the wiggler, without lasing (left image: full momentum spread ~1%, or 400 keV) and with lasing (right image: full momentum spread ~5%, or 2 MeV). This is, as well, indicative of the rather large acceptance required of the recirculator. The recirculator momentum compaction is then used to rotate the bunch so that an appropriate phase energy correlation occurs at reinjection.

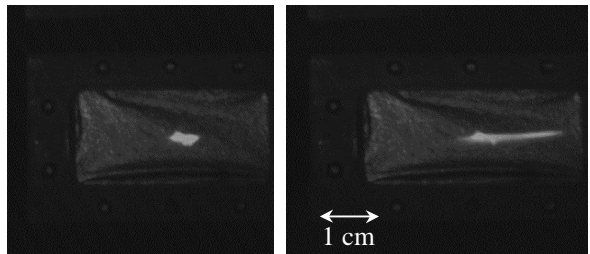

Figure 3: Beam viewer image in downstream chicane (dispersion of 0.4 m); left: no lasing, right: lasing.

The recirculator path length is adjusted by using the aforementioned dipoles to reinject the recirculated beam 180° out of phase with the accelerated beam. This results in a transfer of beam power to the RF structure, with a resulting recovery of the beam energy. The phase-energy correlation imposed by the recirculator trim quads is selected to compensate the slope of the decelerating RF waveform. As a consequence, the 2 MeV energy full spread of the recirculated beam, rather than adiabatically antidamping to a relative energy spread of order 20% during energy recovery to 10 MeV, energy compresses to ~100 keV at the dump, giving a final relative energy spread of 1%. This 20:1 energy compression requires not only the appropriate recirculator $M_{56}$, but also demands the proper $T_{566}$ so as to correct both the lattice quadratic variation of path length with momentum and the curvature of the decelerating RF waveform. Figure 4 illustrates this point by displaying the 10 MeV energy recovered beam at a dispersed point ($\eta$~1 m) near the dump with and without lasing and without and with sextupoles. The beam is more diffuse and the momentum spread greater without sextupoles (top) than when sextupoles are activated (bottom). The final beam spot is however roughly independent of lasing when sextupoles are activated (bottom left, laser off; bottom right, laser on).

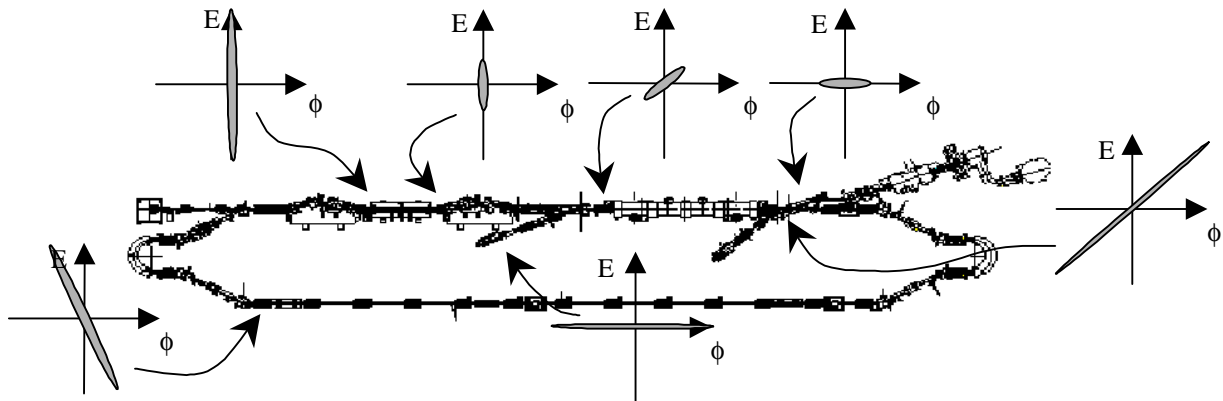

Figure 2: Longitudinal matching scenario in IR Demo showing phase/energy plots at critical locations.

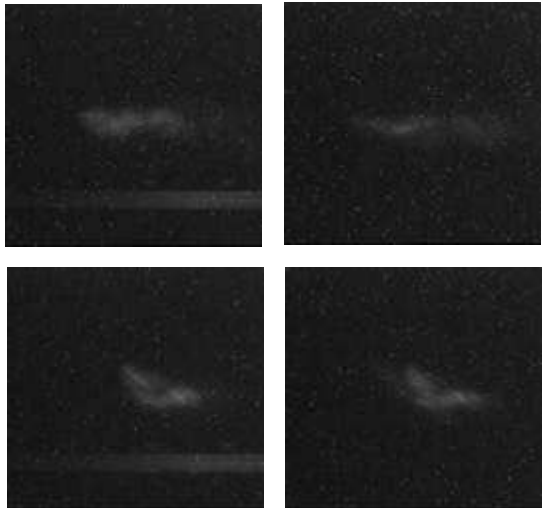

Figure 4: Electron beam spot at dump: top row, sextupoles off; bottom row, sextupoles on; left column, lasing off; right column, lasing on.

Energy recovery is quite efficient. This is illustrated by Figure 5, which presents the RF drive system forward powers in each cryomodule cavity with beam off, with 1 mA of beam without energy recovery, and at various currents with energy recovery. Essentially all of the beam power is recovered, inasmuch as no power beyond the zero current value is required.

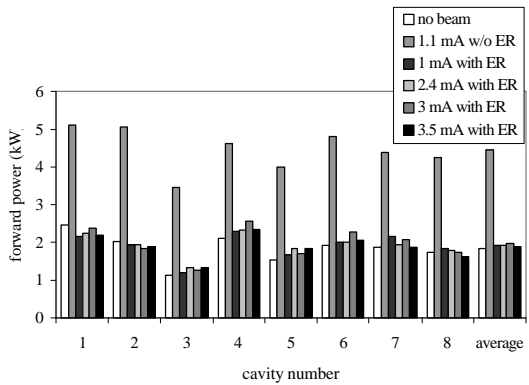

Figure 5: RF drive system forward powers for each cavity without beam, and without and with energy recovery at various current levels.

## 3 CONSTRUCTION AND COMMISSIONING HIGHLIGHTS

IR Demo project funding started in April 1996, with construction and installation continuing through August 1998. Commissioning activities interleaved with construction began in the fall of 1997, with milestones met as indicated in the following chronology:

- Oct. 1997    1st beam in vault (injector)
- Dec. 1997    1st beam to straight-ahead dump
- Mar. 1998    high current single pass operation (1.1 mA CW to straight-ahead dump)
- Jun. 1998    wiggler installed, 1st light (155 W CW at 5 µm /1.1 mA straight ahead)
- Jul. 1998    recirculator construction completed, 1st energy recovered beam, 1st (low power) lasing with energy recovery
- Dec. 1998    high power lasing with energy recovery (>200 W CW at 5 µm/1.4 mA)
- Mar. 1999    kW-class 5 µm operation (710 W CW at 3.6 mA; mirror limited)
- Jul. 1999    1.72 kW CW at 3 µm/4.4 mA; kW-class tuneable light at 3, 5 and 6 µm 5th harmonic (1 µm) lasing
- Sept. 1999   Thomson scattering x-ray production

Early in commissioning, the system was limited to ~30% availability by the gun. Effort in this area has led to a very reliable electron source with nearly 100% availability. The presently installed GaAs wafer has provided cathode lifetimes in excess of 600 C and has delivered over 2 kC total charge [11]. Also noteworthy during commissioning were the production of 1 µm light through fifth harmonic lasing [12] and the generation of intense, short x-ray pulses through Thomson scattering [13]. The latter holds promise of expanding the scope of the user facility to support pump-probe experimentation.

## 4 PERFORMANCE

The driver accelerator and FEL perform flexibly, robustly, and reproducibly. The system restores to full power lasing in a shift after long shutdowns; during normal operations, lasing is recovered in minutes after a vault acess. Operations are simplified by a full suite of diagnostics [14], including beam position monitors, optical transition radiation based beam viewers, beam current montoring cavities and a "Happek device", an interferometric coherent transition radiation based bunch length diagnostic. The former pair of diagnostics allows beam steering and transverse matching, the latter pair supports the longitudinal matching detailed above.

The FEL provides pulsed and CW lasing with variable timing (within limits dictated by the drive laser fundamental of 75 MHz and the optical cavity length of 8 m) over continuously tuneable ranges around 3, 5, and 6 µm (defined by mirror reflectivities). It is used as a source by a growing user community [15] and for machine studies. The latter include the topics of FEL/RF stability, BBU, CSR, and investigations of tapered wiggler dynamics [16]. A typical FEL output spectrum is shown in Figure 6, a detuning curve is shown in Figure 7.

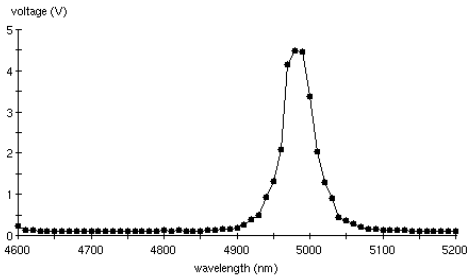

Figure 6: Typical FEL output spectrum.

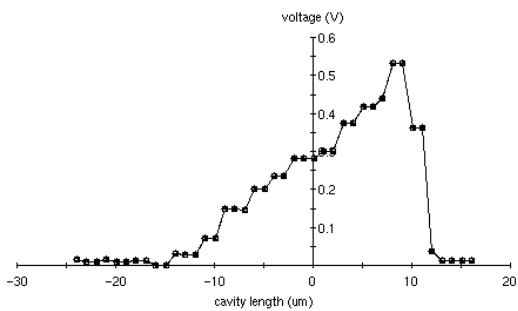

Figure 7: Typical FEL detuning curve.

## 5 THE 10 KW UPGRADE

The U.S. Navy has provided 9.3 M$ initial funding for an upgrade of this system to 10 kW. The envisioned upgrade path will entail
- doubling the injected current from 5 to 10 mA by increasing the bunch charge from 67 to 135 pC,
- installation of two additional cryomodules to raise the beam energy to ~160 MeV,
- upgrading the recirculator to accommodate higher beam energy and a new FEL insertion embedded in the machine backleg, and
- implementation of a 32 m long $R^5$ optical cavity [17] accommodating high power operation on a 2-10 µm bandwidth.

Design, prototyping, and procurement activities are now underway; with anticipated follow-on funding upgrade commissioning is expected to commence in fall 2002. A description of the machine is available elsewhere [18].

## 6 ACKNOWLEDGEMENTS

This paper presents the work of members of the Jefferson Lab IR Demonstration FEL Project Team; I would like to thank them for their efforts and assistance in its preparation. I would also like to thank Dr. Jay Benesch for a careful reading of the text and many useful suggestions. This work was supported by the Office of Naval Research, the Commonwealth of Virginia, and the U.S. Department of Energy through contract number DE-AC05-84ER40150.

## REFERENCES

[1] G. Neil et al., *Phys. Rev. Lett.*, 84, 4:662-5 (2000).

[2] H. Liu et al., *Nuc. Inst. Meth.* A358: 475-8 (1995).

[3] L. Merminga et al., PAC'99, pp. 1177-9, New York, 29 March-2 April 1999.

[4] R. Li et al., PAC'97, Vancouver, May 1997; R. Li, EPAC'98, Stockholm, June 1998; R. Li., FEL'99, Hamburg, August 1999; R Li, EPAC'2000, Vienna, June 2000.

[5] L. Merminga et al., "FEL-RF Instabilities in Recirculating, Energy-Recovering Linacs with an FEL", FEL'99, Hamburg, August 1999.

[6] D. Engwall et al., PAC'97, pp. 2693-5, Vancouver, May 1997.

[7] P. Piot et al., EPAC'98, pp. 1447-9, Stockholm, June 1998; P. Piot et al., "Emittances and Energy Spread Studies in the Jefferson Lab Free-Electron Laser", EPAC'2000, Vienna, June 2000.

[8] H. Liu, private communication; B. Yunn, unpublished.

[9] J. Flanz et al., *Nuc. Inst. Meth.* A241:325-33 (1985).

[10] L. Merminga et al., PAC'99, pp. 2456-8, New York, 29 March-2 April 1999; P. Piot et al., "Study of the Energy Compression Scheme to Energy Recover an Electron Beam in Presence of an FEL Interaction", EPAC'2000, Vienna, June 2000.

[11] T. Siggins et al., "Performance of the Photocathode Gun for the TJNAF FEL", FEL'2000, Durham, N.C., August 2000.

[12] E. Gillman et al., FEL'99, Hamburg, August 1999.

[13] G. Krafft, PAC'99, pp. 2448-9, New York, 29 March-2 April 1999.

[14] G. Krafft et al., PAC'99, pp. 2229-31, New York, 29 March-2 April 1999.

[15] Jefferson Lab IR FEL user facility information is available on-line at http://www.jlab.org/FEL/.

[16] S. Benson et al., "An Experimental Study of an FEL Oscillator with a Linear Taper", FEL'2000, Durham, N.C., August 2000.

[17] C. C. Shih et al., *Nuc. Inst. Meth.* A304:788 (1991).

[18] D. Douglas et al., "Driver Accelerator Design for the 10 kW Upgrade of the Jefferson Lab IR FEL", these proceedings.